\documentclass[10pt,letterpaper,twocolumn]{article}
\usepackage{ol2}
\usepackage[draft,implicit=false]{hyperref}
\usepackage{amsmath}

\begin{document}

\twocolumn[
\title{Tunable deep-subwavelength superscattering using graphene monolayers}

\author{R. J. Li,$^{1,2,3}$ X. Lin,$^{1,2,3}$ S. S. Lin,$^{1,2,3}$  X. Liu,$^{1}$
and H. S. Chen$^{1,2,3,*}$}

\address{
$^1$State Key Laboratory of Modern Optical Instrumentation, Zhejiang University, Hangzhou 310027, China
\\
$^2$Department of Information Science and Electronic Engineering, Zhejiang University, Hangzhou 310027, China
\\
$^3$The Electromagnetics Academy of Zhejiang University, Zhejiang University, Hangzhou 310027, China
\\
$^*$Corresponding author: hansomchen@zju.edu.cn
}

\begin{abstract}
In this Letter, we theoretically propose for the first time that graphene monolayers
can be used for superscatterer designs. We show that the scattering cross section of the bare
deep-subwavelength dielectric cylinder is markedly enhanced by six orders
of magnitude due to the excitation
of the first-order resonance of graphene plamons. By utilizing the tunability of the
plasmonic resonance through tuning graphene's chemical potential, the graphene
superscatterer works in a wide range of frequencies from several terahertz to tens
of terahertz.
\end{abstract}


]

\noindent Subwavelength structures can demonstrate very unusual
electromagnetic properties with the concept of metamaterials \cite%
{nmat11-917}. The scattering of subwavelength structures can be suppressed
to realize various devices, such as invisibility cloaks \cite%
{PRE72-016623,PRL103-153901, PRL102-233901,AM24-OP281,ACSNANO,science-china}%
, but can be also enhanced to realize superscatterers, a kind of device that
can magnify the scattering cross section of a given object remarkably
\cite{OE16-18545} and that has
potential applications ranging from detection, spectroscopy to photovoltaics
\cite{science275-1102,PNAS,JPCC,APL73-3815,NL8-3983,prl90-057401,JAP101-093105,
nmat9-205}.

Transformation optics and complementary metamaterials have been proposed to
dramatically enhance the scattering cross section of a particle \cite{OE16-18545,
NJP11-073033,JOSAA30-1698}.
To implement this method, it requires both electrical and magnetic anisotropic
inhomogeneous parameters. In order to loose the requirement, Ruan et al.
\cite{PRL105-013901,APL98-043101} and Mirzaei et al. \cite{OE21-10454,APL105-011109}
use isotropic plasmonic structures to enhance the scattering cross sections. As the
scattering cross sections of subwavelength structures experience the single channel
limit \cite{PRL97-263902,PRL105-013901}, they use multilayered plasmonic-dielectric
structures to break the single channel limit by engineering an overlap of resonances
of different plasmonic modes.
However, when the sizes of subwavelength structures scale down to the
deep-subwavelength, the scattering cross sections are extremely small and
far below the single channel limit. The original structures in Refs. \cite%
{PRL105-013901,APL98-043101,OE21-10454,APL105-011109} can not be scaled down
directly for superscattering purpose since the commonly available metals in
nanophotonics have a thickness of several nanometers. This naturally raises
the question of how to design a superscatterer which is suitable for the
deep-subwavelength objects.

Graphene, a two dimensional hexagonal crystal carbon sheet with only one atom
thick, can be a good candidate to solve this bottleneck. Since its
ballistic transport and ultrahigh electron mobility, the surface
conductivity of graphene is almost purely imaginary in the THz frequencies
\cite{science332-1291}. In other words, graphene can be treated as a thin
film of metal with low loss. In the recent years, graphene has attracted
much attention as a good counterpart in the THz frequencies of metals in the
optical frequencies, and the research area of graphene plasmonics has
been flourished \cite{LPR,nphoton4-611,nphoton6-749,OE19-6616,PRB89-035406,LPR7-L7,
LPR8-291,OE22-30108}.

In this paper, we theoretically propose for the first time that graphene
monolayers are
used to design the superscatterer which can enlarge the scattering cross
sections of deep-subwavelength dielectric objects in the THz frequencies. We
introduce the model of superscatterer from Mie scattering theory. With the
graphene monolayer, the scattering cross section can be enhanced by six
orders of magnitude. The applicability for dielectric media with different
permittivities and different incident frequencies are analysed by utilizing
the tunability of surface conductivity of graphene.

\begin{figure}[tbp]
\centering
\vspace{-0.5cm}
\centerline{\includegraphics[width=8cm]{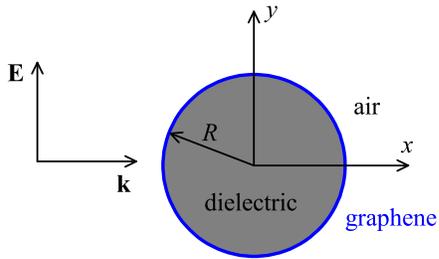}}
\vspace{-1.5cm}
\caption{Cross-sectional view of the superscatterer. The cylindrical dielectric
medium (grey area) is coated with the graphene monolayer (blue area). A
TM-polarized plane wave with magnitude $H_0$ normally incidents onto the
superscatterer. The radius of the dielectric medium is $R$.}
\label{superscatter}
\end{figure}

As shown in Fig. \ref{superscatter}, the case of a TM-polarized plane wave
with magnitude $H_{0}$ normally incidents from air onto an infinite long
graphene coated
cylindrical dielectric medium is considered. The incident magnetic field is $%
\mathbf{H}=H_{0}e^{ik_{0}x}\hat{z}$ with the time dependence of $\exp \left(
-i\omega t\right) $, where $k_{0}=\omega \sqrt{\varepsilon _{0}\mu _{0}}$ is
the wavenumber in free space and $\omega $ is the angular frequency of the
incident field. The radius of the dielectric medium is $R$. The relative
permittivity of the dielectric medium is $\varepsilon _{r}$, and the
relative permeability is $\mu _{r}=1$.

Since graphene is a two dimensional electromagnetic material and its
thickness is extremely small compared with the radius of the cylindrical
dielectric medium, we will treat it as a conducting film with surface
conductivity $\sigma _{g}$ \cite{science332-1291,JAP103-064302}. From Mie
scattering theory, the magnetic field can be written as $H_{z}=H_{0}\sum%
\nolimits_{n=0}^{\infty }a_{n}\left[ J_{n}\left( k_{0}r\right)
+s_{n}H_{n}^{\left( 1\right) }\left( k_{0}r\right) \right] \cos \left(
n\theta \right) $ for $r>R$, and $H_{z}=H_{0}\sum\nolimits_{n=0}^{\infty
}b_{n}J_{n}\left( kr\right) \cos \left( n\theta \right) $ for $r\leq R$,
where $a_{n}=\delta _{n}i^{n}$ ($\delta _{n}=1$ for $n=0$ and $\delta _{n}=2$
for $n\neq 0$), $k=k_{0}\sqrt{\varepsilon _{r}}$ is the wavenumber in
dielectric medium, $s_{n}$ is the scattering coefficient, $J_{n}$ and $%
H_{n}^{\left( 1\right) }$ are the $n$-th order Bessel function of the first
kind and Hankel function of the first kind, respectively \cite{book2}.
According to the continuity conditions at $r=R$, the scattering coefficient
can be obtained as%
\begin{equation}
s_{n}=-\frac{J_{n}^{\prime }\left( k_{0}R\right) t_{n}-J_{n}\left(
k_{0}R\right) J_{n}^{\prime }\left( kR\right) }{H_{n}^{\left( 1\right)
\prime }\left( k_{0}R\right) t_{n}-H_{n}^{\left( 1\right) }\left(
k_{0}R\right) J_{n}^{\prime }\left( kR\right) },  \label{1}
\end{equation}%
where $t_{n}=\sqrt{\varepsilon _{r}}J_{n}\left( kR\right) +i\sigma _{g}\eta
_{0}J_{n}^{\prime }\left( kR\right) $, $\sigma _{g}$ is the surface
conductivity of graphene, $\eta _{0}=\sqrt{\mu _{0}/\varepsilon _{0}}$ is
the impedance of free space. We define the normalized scattering cross
section (NSCS) as NSCS$=\sum\nolimits_{n=0}^{\infty }\delta _{n}\left\vert
s_{n}\right\vert ^{2}$, which is normalized by $2\lambda /\pi $ and we have
considered the degeneracy between $\left\vert s_{n}\right\vert $ and $%
\left\vert s_{-n}\right\vert $ \cite{APL105-011109}. Note when $\sigma
_{g}=0 $, it reduces to the case of scattering by a bare cylindrical
dielectric medium.

The surface conductivity of graphene can be calculated according to the Kubo
formula $\sigma _{g}\left( \omega ,\mu _{c},\Gamma ,T\right) =\sigma
_{intra}+\sigma _{inter}$, where
\small
\begin{equation}
\sigma _{intra}=\frac{ie^{2}k_{B}T}{\pi \hbar ^{2}\left( \omega +i2\Gamma
\right) }\left[ \frac{\mu _{c}}{k_{B}T}+2\ln \left( e^{-\mu
_{c}/k_{B}T}+1\right) \right]   \label{2}
\end{equation}%
\normalsize
is due to intraband contribution, and
\small
\begin{equation}
\sigma _{inter}=\frac{ie^{2}\left( \omega +i2\Gamma \right) }{\pi \hbar ^{2}}%
\int_{0}^{\infty }\frac{f_{d}\left( -\varepsilon \right) -f_{d}\left(
\varepsilon \right) }{\left( \omega +i2\Gamma \right) ^{2}-4\left(
\varepsilon /\hbar \right) ^{2}}d\varepsilon   \label{3}
\end{equation}%
\normalsize
is due to interband contribution \cite%
{JAP103-064302,JP19-026222}. In the above formula, $-e$ is the charge of an
electron, $\hbar =h/2\pi $ is the reduced Plank's constant, $\omega =2\pi f$
is the angular frequency of the incident field, $\Gamma $ is the
phenomenological scattering rate that is assumed to be independent of the
energy $\varepsilon $, $f_{d}\left( \varepsilon \right) =1/\left( e^{\left(
\varepsilon -\mu _{c}\right) /k_{B}T}+1\right) $ is the Fermi-Dirac
distribution, $k_{B}$ is the Boltzmann's constant, $T$ is the temperature,
and $\mu _{c}$ is the chemical potential which can be tuned by a gate
voltage and/or chemical doping. In the following, we choose $\Gamma =0.11$
meV, $T=300$ K and the chemical potential $\mu _{c}$ is tuned between $0$ eV
to $0.5$ eV \cite{JAP103-064302}. Thus, for a given frequency and chemical potential we can
obtain the surface conductivity of graphene and calculate the NSCS of the
superscatterer for a given inside dielectric medium.
\begin{figure}[tbp]
\centering
\vspace{0.0cm}
\centerline{\includegraphics[width=8cm]{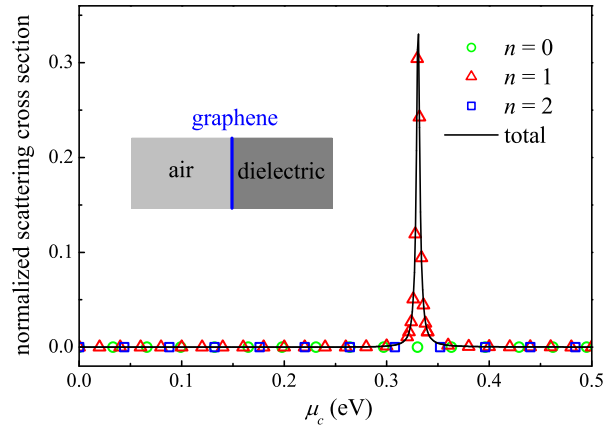}}
\vspace{-0.1cm}
\caption{Schematically shows the total normalized scattering cross section
(NSCS) and the contributions of the first three scattering terms as a
function of the chemical potential $\protect\mu _{c}$. The inset shows the
equivalent planar structure. The parameters are $\protect\varepsilon %
_{r}=1.44$, $f=15$ THz and $R=0.2$ $\protect\mu $m.}
\label{resonance}
\end{figure}
As an example, we let $\varepsilon _{r}=1.44$, $f=15$ THz and $R=0.2$ $\mu m$%
. Under these parameters, this structure corresponds to a graphene coated
deep-subwavelength dielectric cylinder. Fig. \ref{resonance} schematically
shows the dependence between NSCS and the chemical potential $\mu _{c}$
which exhibits a sharp resonance. The superscattering occurs at $\mu
_{c}=0.331$ eV with the maximum NSCS equals to $0.343$. For comparison, we
also plot the contributions of the first three scattering terms in the
figure. Clearly, the resonance of NSCS is caused by the resonance of the
first order scattering term ($n=1$). To validate this, we use the series
forms of Bessel function and Hankel function to simplify the scattering
coefficient $s_{n}$ \cite{book3}. For a deep-subwavelength dielectric
cylinder which satisfies $kR\ll 1$, only the first orders of the series are
needed. Detailed calculations show%
\begin{equation}
s_{1}=\frac{i\pi k_{0}^{2}R^{2}}{4}\frac{\left( \varepsilon -1\right)
k_{0}R+i\sigma _{g}\eta _{0}}{\left( \varepsilon +1\right) k_{0}R+i\sigma
_{g}\eta _{0}},  \label{4}
\end{equation}%
which exhibits a resonance at $\sigma _{g}=i\left( \varepsilon +1\right)
k_{0}R/\eta _{0}=0.407i$ mS when the graphene is assumed to be lossless. It is
approximately equal to the result from Mie scattering theory where $\sigma
_{g}=0.001+0.409i$ mS. Since the real part of conductivity is small compared
with its imaginary part, this superscatterer has a large scattering cross
section with low energy dissipation. Note when there is no graphene, the
NSCS of the dielectric medium is $6.254\times 10^{-7}$. This indicates that
coating with the properly doped and/or gated graphene monolayer can greatly
enhance the scattering by six orders of magnitude. Although this kind of
superscatterer is realized by only one resonant mode which still
experiences the single channel limit and it is different to Refs. \cite%
{PRL105-013901,APL98-043101, OE21-10454,APL105-011109}, the scattering cross
section has enlarged significantly which is enough to demonstrate the
superscattering phenomenon.

\begin{figure}[tbp]
\vspace{-0cm}
\centerline{\includegraphics[width=8cm]{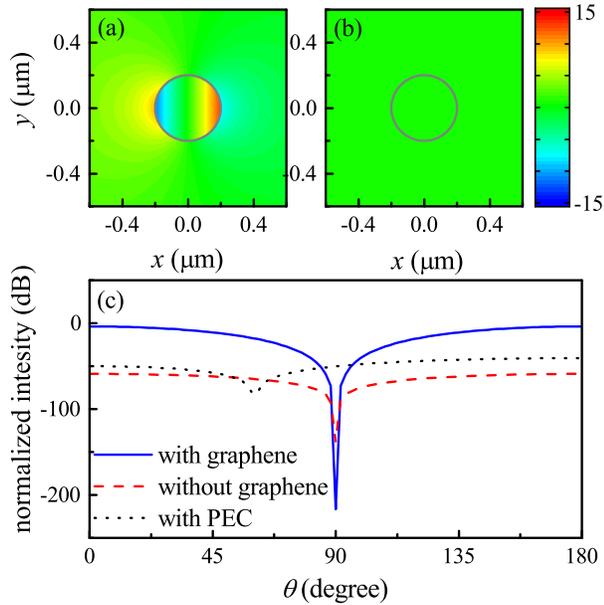}}
\vspace{-0cm}
\caption{Top panel: normalized magnetic field distributions for (a)
dielectric medium coated with the graphene monolayer and (b) the bare
dielectric medium. Bottom panel: far field scattering patterns for the
dielectric medium coated with graphene (blue solid line), without graphene
(red dashed line) and with PEC (black dotted line), respectively. The grey
circles in (a) and (b) indicate the boundary of the dielectric cylinder.
The parameters are $\protect\varepsilon %
_{r}=1.44$, $f=15$ THz, $R=0.2$ $\protect\mu $m and $\mu
_{c}=0.331$ eV.}
\label{field}
\end{figure}

There are two critical conditions to realize this kind of
superscattering: permittivity and optical loss, which correspond to the imaginary
part and real part of the surface conductivity, respectively. The
permittivity of the coating layer must be delicately determined to satisfy
the resonant condition \cite{PRL105-013901,arxiv}. The inset in Fig. \ref%
{resonance} shows the equivalent planar structure, and its corresponding
dispersion relation of the graphene plamon is
\begin{equation}
\left( 1+i\frac{\sigma _{g}k _{1}}{\omega \varepsilon _{0}\varepsilon
_{r}}\right)k_{2}+\frac{k_{1}}{\varepsilon _{r}}=0,  \label{5}
\end{equation}%
where $k_{1}=\left( \beta ^{2}-k_{0}^{2}\varepsilon _{r}\right) ^{1/2}$%
, $k _{2}=\left( \beta ^{2}-k_{0}^{2}\right) ^{1/2}$, and $\beta $ is
the propagation constant. For parameters in Fig. \ref{resonance} and
assuming that the graphene is lossless, the propagation constant is $\beta
=15.878k_{0}$ at $\mu _{c}=0.331$ eV which is nearly equal to the first
order Bohr condition $\beta =50/\pi k_{0}$ \cite{arxiv}. Meanwhile, the optical
loss of the
coating layer should be small enough. Actually, if we increase the real part
of surface conductivity of graphene by 10 times, i.e. $\sigma
_{g}=0.010+0.409i$ mS, the corresponding NSCS at the resonant point $\mu
_{c}=0.331$ eV would be 0.008 which is only 2.5\% of the original value.
Moreover, the thickness of graphene is small compared with the radius of the
deep-subwavelength dielectric cylinder. Thus graphene is indeed a good
choice for the study of superscattering of deep-subwavelength dielectric
objects and the design of corresponding superscatterers.

Fig. \ref{field} schematically shows the normalized magnetic
field distributions and far field scattering patterns. When the dielectric
medium is coated with the graphene monolayer with $\mu
_{c}=0.331$ eV, the normalized magnetic field
intensity at the center of the medium is almost zero and the scattering is
enhanced both in the forward direction and the backward direction
(See Fig. \ref{field} (a)) . Whereas, when the
graphene is removed, the corresponding magnetic field is almost unperturbed
as shown in Fig. \ref{field} (b)-(c). Note although zero field intensity
inside the dielectric cylinder can also be realized just by coating the
dielectric medium with a PEC, however, as shown in Fig. \ref{field} (c), the
far field scattering patterns of graphene coated dielectric medium and PEC
coated dielectric medium are totally different. This can be understood by
calculating the equivalent relative permittivity of graphene.

Under our optimized parameters, the permittivity of graphene is
$\varepsilon_g =-489.4+1.8i$ which is calculated by the formula $%
\varepsilon_g =1+i\sigma _{g}/\omega \varepsilon _{0}d$, where $d$ is the
thickness of graphene. In the calculation, we let $d=1$ nm which is commonly
used in the simulation of graphene plasmonics\cite{science332-1291}. Thus graphene can
be treated as a thin film of low loss metal. Whereas, the PEC is a kind of
metal with infinite loss. In fact, calculations show that the NSCS of PEC
coated dielectric medium is $2.883\times 10^{-5}$. This indicates that our
result is not the trivial outcome with the use of negative permittivity
materials.

\begin{figure}[tbp]
\vspace{-0cm}
\centerline{\includegraphics[width=8cm]{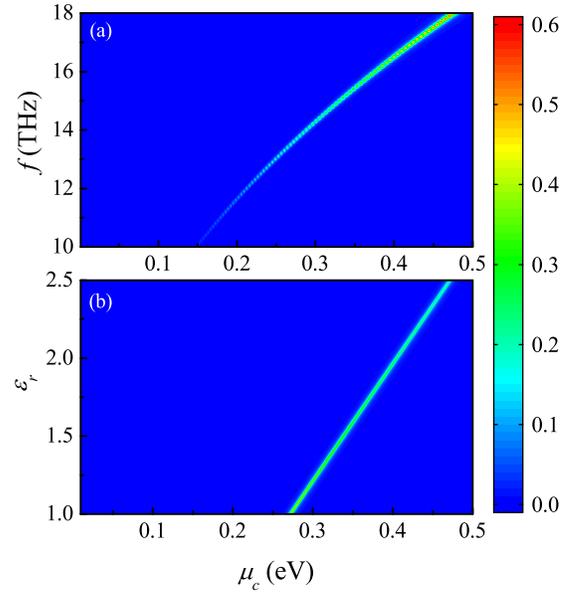}}
\vspace{-0cm}
\caption{The normalized scattering cross sections (NSCSs) for (a) different
incident frequencies and (b) dielectric media with different permittivities
when the chemical potential $\protect\mu_c$ is tuned between 0 eV to 0.5 eV,
respectively. The parameters are $\Gamma =0.11$ meV, $T=300$ K, $R=0.2$ $%
\protect\mu $m for both (a) and (b), while $\protect\varepsilon_r =1.44$ for
(a), and $f=15$ THz for (b).}
\label{tunability}
\end{figure}

In practical applications, the tunability of a device is of
great importance. Since the chemical potential of graphene can be tuned by a
gate voltage and/or chemical doping, this provides a versatile method to
manipulate the superscattering of deep-subwavelength dielectric objects. To
this end, we vary the frequency of incident wave and the permittivity of
dielectric medium, respectively, and calculate the corresponding NSCS when
the chemical potential $\mu_c$ is tuned between 0 eV to 0.5 eV. For
simplicity, the dispersion of the dielectric medium is neglected. As shown
in Fig. \ref{tunability}(a), the superscattering occurs between $f=10$ THz
to $f=18$ THz with relative large values of NSCSs, which indicates that we
can tune the chemical potential of graphene delicately to get the
superscattering phenomenon. Although the maximum NSCS at lower frequencies
is less than 0.04, it is still much larger than the NSCS of the bare
dielectric medium. Similarly, the superscattering also occurs when the
relative permittivity varies between $\varepsilon_r =1$ to $\varepsilon_r
=2.5$ and their corresponding NSCSs have relative large values as shown in
Fig. \ref{tunability}(b). This indicates that this superscatterer is suitable
for dielectric media with low permittivities.

In conclusion, we demonstrate the possibility of
superscattering of deep-subwavelength dielectric objects by graphene
monolayers. The resonance of the first order scattering term can be emerged
by tuning the chemical potential of graphene. This superscatterer is
applicable in a wide range of frequencies from several terahertz to tens of
terahertz and it is suitable for the superscattering of dielectric media
with low permittivities. This superscattering effect may be useful in
practical THz devices such as detectors and sensors.

This work was sponsored by the National Natural Science
Foundation of China under Grants No. 61322501, and No. 61275183, the
National Program for Special Support of Top-Notch Young Professionals, the
Program for New Century Excellent Talents (NCET-12-0489) in University, and
the Fundamental Research Funds for the Central Universities
(2014XZZX003-24).

\pagebreak

\section*{\protect\normalsize Informational Fifth Page}

In this section, please provide full versions of citations to
assist reviewers and editors (OL publishes a short form of citations) or any
other information that would aid the peer-review process.

\end{document}